\documentclass[12pt,preprint]{aastex}

\def\beq{\begin{eqnarray}}
\def\eeq{\end{eqnarray}}
\def\d{{\rm d}}
\def\dz{{\rm d}z}
\def\dln{{\rm dln}}
\def\p{\partial}
\def\a{\alpha}
\def\rg{r_{\rm g}}

\def\vr{v_r}
\def\cs{c_{\rm s}}
\def\Om{\Omega}
\def\OmK{\Omega_{\rm K}}
\def\dM{\dot M}
\def\dm{\dot m}

\def\Qvis{Q_{\rm vis}^+}
\def\Qadv{Q_{\rm adv}^-}
\def\Qrad{Q_{\rm rad}^-}
\def\fadv{f_{\rm adv}}
\def\dmax{\dot m_{\rm max}}
\def\kes{\kappa_{\rm es}}
\def\MBH{M_{\rm BH}}

\shorttitle{Slim Disks}
\shortauthors{Gu \& Lu}
\shortauthors{}

\begin{document}

\title{A Note on the Slim Accretion Disk Model}

\author{Wei-Min Gu and Ju-Fu Lu}

\affil{Department of Physics
and Institute of Theoretical Physics and Astrophysics, \\
Xiamen University, Xiamen, Fujian 361005, China}

\email{guwm@xmu.edu.cn}

\begin{abstract}
We show that when the gravitational force is correctly calculated
in dealing with the vertical hydrostatic equilibrium of black hole
accretion disks, the relationship that is valid
for geometrically thin disks, i.e., $\cs/\OmK H =$ constant,
where $\cs$ is the sound speed, $\OmK$ is the Keplerian angular velocity,
and $H$ is the half-thickness of the disk, does not hold for slim
disks. More importantly, by adopting the correct vertical gravitational
force in studies of thermal equilibrium solutions,
we find that there exists a maximally possible
accretion rate for each radius in the outer region of optically thick
accretion flows, so that only the inner region of these flows can
possibly take the form of slim disks, and strong outflows from the outer
region are required to reduce the accretion rate in order for slim disks
to be realized.
\end{abstract}


\section{Introduction}

The slim disk model is one of popular models for accretion flows
around black holes and has been applied in recent years to quite
many energetic astrophysical systems such as narrow line
Seyfert 1 galaxies (e.g., Wang \& Netzer 2003; Chen \& Wang 2004),
Galactic black hole candidates (e.g., Watarai et al. 2000),
and ultraluminous X-ray sources
(e.g., Watarai, Mizuno, \& Mineshige 2001; Vierdaynti et al. 2006).

Despite of its growing importance in the observational sense, we notice
that some theoretical problems regarding to this model in the
fundamental sense have been ignored. In this paper we address one
such problem, namely the inaccurate calculation of
the gravitational force in dealing with the hydrostatic equilibrium
in the vertical direction of slim disks.
We work in the cylindrical coordinate system ($r$, $z$, $\varphi$) throughout.

\section{Vertical gravitational force}

Obviously because the Shakura-Sunyaev disk (SSD) model
(Shakura \& Sunyaev 1973) is the first and still the most successful
model of accretion disks, many simplifications made in this model are
followed by the subsequent accretion disk models without rigorously
verifying their applicability. The treatment of the disk's vertical
structure is such a one.

The very basic character of an SSD is that it is geometrically thin,
i.e., everywhere in the disk the half thickness $H(r)$ is much smaller
than the corresponding cylindrical radius $r$,  $H(r)/r \ll 1$.
This means that the averaged motion of disk matter in the vertical
direction (if there is any) must be negligible comparing with that
in the radial direction, and it is reasonable to assume that
in the macro sense the disk matter is in the vertical
hydrostatic equilibrium, i.e., in the vertical direction the gravitational
force and the pressure force are balanced with each other,
\beq
\frac{\p p}{\p z} + \rho \frac{\p \psi}{\p z} = 0 \ ,
\eeq
where $p$ is the pressure, $\rho$ is the mass density, $\psi$ is
the gravitational potential, and the disk is already assumed to be
steady and axisymmetric, $\p/\p t = \p/\p \varphi = 0$.
In the SSD model the Newtonian potential
$\psi_{\rm N} (r,z) = -GM /\sqrt{r^2+z^2}$
was used, where $M$ is the mass of central accreting object,
and the one-zone approximation in the vertical direction was made;
then in equation~(1) $\p p/\p z \simeq -p/H$,
$\p \psi_{\rm N} /\p z \simeq GMH/r^3 = \OmK^2 H$,
where $\OmK = (GM/r^3)^{1/2}$ is the Keplerian angular velocity,
and a relation $H = \cs/\OmK$ was obtained, where $\cs = (p/\rho)^{1/2}$
is the sound speed (e.g., Kato, Fukue, \& Mineshige 1998, p.~80).

The above procedure was somewhat improved by H\={o}shi (1977),
who approximated the Newtonian potential as
$\psi_{\rm N} (r,z) \simeq \psi_{\rm N} (r,0) + \OmK^2 z^2/2$
and assumed a polytropic relation in the vertical direction,
$p = K\rho^{1+1/N}$, where $K$ and $N$ are constants,
instead of the one-zone approximation.
Then the disk becomes to have a vertical structure, and the
vertical integration of equation~(1) gives $H = \sqrt{2(N+1)}\cs/\OmK$,
where the sound speed $\cs = (p_0/\rho_0)^{1/2}$,
with subscript 0 representing quantities on the equatorial plane.

Both the SSD model and H\={o}shi (1977) reduced the differential
equation~(1) into a very simple but very useful relation:
$\cs/\OmK H =$ constant. This relation was adopted in the
slim disk model in the following way.
First, the vertical hydrostatic equilibrium was still assumed,
even though slim disks are not geometrically thin, i.e.,
they may have $H \la r$ or $H \sim r$.
The argument for this was that these disks have quite large
radial velocities, so that their configurations are quasi-spherical,
i.e., the vertical motion of disk matter can still be safely neglected,
and equation~(1) holds. Second, although the pseudo-Newtonian
potential introduced by Paczy\'{n}ski \& Wiita (1980,
hereafter the PW potential), i.e.,
\beq
\psi_{\rm PW} (r,z) = -G\MBH /(\sqrt{r^2+z^2}-\rg) \ ,
\eeq
was widely used to simulate the general relativistic effect of
central black hole, where $\MBH$ is the black hole mass and
$\rg \equiv 2G\MBH/c^2$ is the Schwarzschild radius, it was
again treated in the form of H\={o}shi (1977), i.e.,
\beq
\psi_{\rm PW} (r,z) \simeq \psi_{\rm PW} (r,0) + \OmK^2 z^2/2 \ ,
\eeq
where the Keplerian angular velocity $\OmK = (GM/r)^{1/2}/(r-\rg)$.
Using equation (3) and keeping the assumption of polytropic relation,
the relation $\cs/\OmK H =$ constant was refined by integrating
equation (1),
with the sound speed $\cs$ being defined either
in terms of the pressure and mass density on the equatorial
plane (Abramowicz et al. 1988; Wang \& Zhou 1999; Chen \& Wang 2004),
or in terms of the vertically integrated pressure and density,
i.e., $\cs = (\Pi/\Sigma)^{1/2}$, where $\Pi = \int_{-H}^{H} p \dz$
and $\Sigma = \int_{-H}^{H} \rho \dz$
(Kato et al. 1998, p.~242; Watarai 2006).

To check the validity of the H\={o}shi form of potential, we draw
in Figure~1 the gravitational force in the vertical direction
$\p \psi_{\rm PW} / \p z$ in units of ($c^2/\rg$) for varying
values of ($z/r$) at a fixed radius $r = 10\rg$, calculated from
equations (2) (the solid line) and (3) (the dashed line), respectively.
It is as expected that the H\={o}shi form of potential equation~(3)
is valid only for $(z/r) \la 0.2$, while for $(z/r) \sim 1$ it greatly
magnifies the vertical gravitational force comparing with the correct
result according to the explicit form of the PW potential equation~(2).

In the following calculations we take the polytropic index $N=3$.
Instead of the simple relationships
$(\rho/\rho_0)^{1/3} = (p/p_0)^{1/4} = 1-(z/H)^2$ and
$8p_0/\rho_0 = \OmK^2 H^2$ obtained using equation (3)
(Kato et al. 1998, p.~241),
the vertical integration of equation~(1) using equation~(2) gives
\beq
\left( \frac{\rho}{\rho_0} \right) ^{1/3}
= \left( \frac{p}{p_0} \right) ^{1/4}
= \frac{\frac{1}{\sqrt{1+(z/r)^2}-\rg/r} - \frac{1}{\sqrt{1+(H/r)^2}-\rg/r}}
{\frac{1}{1-\rg/r}-\frac{1}{\sqrt{1+(H/r)^2}-\rg/r}} \ ,
\eeq
\beq
4 \frac{p_0}{\rho_0} = \OmK^2 r^2 (1-\rg/r)
\left[ 1 - \frac{1-\rg/r}{\sqrt{1+(H/r)^2}-\rg/r} \right] \ .
\eeq
For a fixed $r$ and a given value of $H/r$, we obtain values of
($\Sigma/2\rho_0 H$) and ($\Pi/2p_0 H$) by vertically integrating
equation~(4), and value of ($\cs/\OmK H$) by using equation~(5).
We define $\cs^2 = \Pi/\Sigma$ as in Kato et al. (1998),
where detailed results were provided and can be quantitatively
compared with ours.

Figure~2 shows $\Sigma/2\rho_0 H$ (the dashed line), $\Pi/2p_0 H$
(the dotted line), and $\cs/\OmK H$ (the solid line) as functions of
$H/r$ at $r=10\rg$. In the slim disk model these three quantities
were all constant, for $N=3$ they were: $\Sigma/2\rho_0 H = 16/35$,
$\Pi/2p_0 H = 128/315$, and $\cs/\OmK H = 1/3$
(Kato et al. 1998, p.~242).
It is seen from the figure that, however, these quantities are not
constant, they take approximately their model-predicted values
only for $H/r \ll 1$. For slim disks with $H/r \sim 1$, the
simple relation $\cs/\OmK H =$ constant is invalid.

\section{Thermal equilibria}

To reveal further the consequence of the correction of vertical
gravitational force, we go on to study
thermal equilibrium solutions of black hole accretion flows.
We write the continuity,
radial momentum, angular momentum, energy, and state equations
of accretion flows in the form similar to that of Kato et al. (1998):
\beq
\dM = -2 \pi r \Sigma \vr = {\rm const.} \ ,
\eeq
\beq
\vr\frac{\d \vr}{\dln r} + \cs^2 \frac{\dln \Pi}{\dln r} - \Om^2 r^2
= - \frac{\OmK^2 (r-\rg)^2}{\Sigma} \int_{-H}^{H}
\frac{\rho}{[\sqrt{1+(z/r)^2}-\rg/r]^2 \sqrt{1+(z/r)^2}} \d z \ ,
\eeq
\beq
\dM (\Om r^2-j_0) = 2 \pi \a r^2 \Pi \ ,
\eeq
\beq
\Qvis = \Qadv + \Qrad \ ,
\eeq
\beq
\Pi = \Pi_{\rm gas} + \Pi_{\rm rad}
= \frac{k_{\rm B} \rho_0 T_0}{\mu m_{\rm p}} \int_{-H}^{H}
\left( \frac{\rho}{\rho_0} \right) \left( \frac{T}{T_0} \right) \dz
+ \frac{1}{3} a T_0^4 \int_{-H}^{H}
\left( \frac{T}{T_0} \right) ^4 \dz \ ,
\eeq
where $\dM$ is the mass accretion rate, $\vr$ is the radial velocity,
$\Om$ is the angular velocity,
$j_0$ is an integration constant representing the
specific angular momentum (per unit mass) accreted by the
black hole, and $\a$ is the viscosity parameter.
Equation~(7) is obtained by vertically integrating equation~(8.2)
of Kato et al. (1998)
and specifying the PW potential equation (2). The viscous heating rate is
$\Qvis = \dM \Om^2 f g / 2\pi$, where $f = 1-j/\OmK r^2$,
$j=j_0 /\omega$, $\omega = \Om/\OmK$ is assumed to be a constant
that is smaller than 1 (sub-Keplerian rotation) and is to be evaluated,
and $g = -\dln \OmK/\dln r$.
The advective cooling rate is
$\Qadv = \xi \dM \cs^2 / 2\pi r^2$
(see Kato et al. 1998 for the detailed expression of $\xi$).
The radiative cooling rate is $\Qrad = 32 \sigma T_0^4 / 3 \tau$,
where $T_0$ is the temperature on the equatorial plane,
$\tau = \kes \rho_0 H$ is the vertical optical depth
(e.g., Eq.~[8.61] of Kato et al. 1998),
and $\kes \simeq 0.34$ cm$^2$g$^{-1}$ is the
electron scattering opacity that is assumed to be the
dominant opacity source.
In equation~(10) $T$ is the temperature,
and $T/T_0 = (\rho/\rho_0)^{1/3}$ as expressed by equation~(4).
For slim disks the radiation pressure dominates over the
gas pressure, so the term $\Pi_{\rm gas}$ in equation~(10) can
be dropped for the moment.

To avoid unnecessary complicacy, we ignore the ram pressure term
$\vr\d \vr / \dln r$ in equation~(7), and take $\dln \Pi / \dln r = -3/2$
and $\xi = 3/2$ from the self-similar solution (e.g., Wang \& Zhou 1999;
Watarai 2006). Then we have a set of six algebraic equations,
i.e., equations~(5-10), which can be solved for six unknown quantities
$H$, $\vr$, $\cs$, $\Om$ (or $\omega$), $\rho_0$, and $T_0$,
with given values of $\MBH$, $\dM$, $\alpha$, $j$, and $r$.
In our calculations we fix $\MBH = 10 M_{\sun}$, $\alpha = 0.1$,
and $j=1.83 c\rg$
(a reasonable value that is just a little less than the Keplerian
angular momentum at the last stable orbit, $\OmK r^2|_{3\rg} = 1.837 c\rg$).

Figure~3 shows thermal equilibrium solutions at a certain radius
in the form of $\dm-H/r$ plane, where $\dm$ is the accretion
rate normalized by the Eddington accretion rate
$\dot M_{\rm Edd} = 64 \pi G\MBH/c\kes$.
The thick lines represent the solutions obtained with the PW potential
equation (2). For comparison, the thin lines draw the solutions obtained with
the H\={o}shi form of potential equation~(3), and accordingly
equation~(7) is reduced to equation~(2.9) of Matsumoto et al. (1984).
The dashed, solid, and dotted lines are for
$r=5\rg$, $r=50\rg$, and $r=500\rg$, respectively.
It is seen that for $r=5\rg$, the difference between the results
obtained with two potentials is only quantitative.
For $r=50\rg$ and $r=500\rg$, however, the difference becomes qualitative.
The most remarkable is that the thick solid and thick dotted lines have
a maximum (not drawn in the figure), i.e.,
there exists a maximally possible
accretion rate $\dmax$ for slim disk solutions in the PW potential,
while no such a $\dmax$ exists for solutions in 
the H\={o}shi form of this potential (the thin solid and thin dotted lines).
One might wonder whether $H/r > 1$, at which $\dmax$ appears,
is physical. Our arguments are the following. First, $H/r > 1$ has
indeed been obtained in many works on slim disks or other accretion
disks (see, e.g.,
Figs.~5-10 of Peitz \& Appl 1997; Figs.~1-4 of Popham \& Gammie 1998;
Figs.~4 and 6 of Lu, Gu, \& Yuan 1999; Fig.~5 of Chen \& Wang 2004;
Figs.~3 and 5 of Watarai 2006). Second, it does not matter for which
value of $H/r$ this $\dmax$ really appears. What is important is that
for each large radius there is an upper limit for $\dm$, beyond which
no thermal equilibrium solutions can be constructed. By contrast, the
previous understanding in the slim disk model (e.g., Abramowicz et al.
1988; Chen et al. 1995; Kato et al. 1998) was that any large $\dm$
can correspond to a thermal equilibrium
solution, as the thin solid and thin dotted lines in Figure~3 imply.
This is because those authors either used the H\={o}shi form of potential
equation~(3) that magnifies the vertical gravitational
force, or used the relation $\cs/\OmK H =$ constant that is resulted
from equation~(3).

It is also noticeable that at small radii ($r = 5\rg$ in Fig.~3)
there is no $\dmax$ for thermal equilibrium solutions. Figure~4 illustrates
this qualitative difference between small and large radii. Figure~4(a)
is for $\dm = 10$ and $r = 5\rg$, Figure~4(b) is also for $\dm = 10$ but
for $r = 50\rg$ ($\dm$ is larger than the corresponding $\dmax$). It is
clear that in the former case the thermal equilibrium can be established,
i.e., $\Qvis = Q^{-}$ ($\equiv \Qadv + \Qrad$), or $\Qvis = \Qadv$ if
$\Qrad$ is totally negligible, can be realized for some value of $H/r \sim 1$;
while in the latter case there are no thermal equilibrium solutions as
$\Qvis$ is always larger than $Q^{-}$.

We derive an approximate analytic expression of the critical radius
$r_{\rm crit}$, for $r > r_{\rm crit}$ there is a $\dmax$ and for
$r < r_{\rm crit}$ there is not. Equation~(5) gives $4\cs^2 < \OmK^2 r^2$,
and equation~(7) gives $(3/2)\cs^2 + \Om^2 r^2 \simeq \OmK^2 r^2$,
so we have $(5/2)\cs^2 < \Om^2 r^2$. As seen from Figure~4,
the condition that can ensure a thermal equilibrium solution
at a radius is $\Qadv \ge \Qvis$,
i.e., $(3/2)\cs^2/r^2 \ge \Om^2 f g \simeq (3/2)\Om^2 f$ since
$g \simeq 3/2$. Therefore, the criterion for a thermal equilibrium solution
to exist is expressed as
\beq
f \equiv 1 - \frac{j}{\OmK r^2} \le \frac{2}{5} \ .
\eeq
For $j = 1.83 c\rg$ in our calculations, the critical value
$f_{\rm crit} = 2/5$ gives $r_{\rm crit} = 16.4 \rg$, which agrees
well with the numerical value $r_{\rm crit} = 18.8 \rg$
in Figures 3 and 5.
The reason why $f$ is of crucial importance is that since $\Qvis$ is
proportional to $f$, for small radii $f$ is small, and $\Qvis$ can
be balanced by $\Qadv$ for any $\dm$; while for large radii $f$
is large, there must be an upper limit for $\dm$, beyond which
$\Qvis$ would be too large to be balanced by any cooling.

To see where the position of slim disks is according to
our understandings, we present in Figure~5 a united description
of thermal equilibrium solutions of optically thick accretion
flows around black holes in the form of $\dm - r$ plane.
The solid line in the figure draws $\dmax$ for each radius,
above which no thermal equilibrium solutions exist at all.
The rest of the plane is further divided into three regions
based on the local stability analysis. We adopt a simple thermal
instability criterion: 
$(\p \Qvis / \p T)_{\Sigma} - (\p Q^{-} / \p T)_{\Sigma} > 0$,
which is practically valid for moderately large-scale perturbations
(Kato et al. 1998, p.~306). The criterion results in: 
$\delta = 2 - 5 \beta - 6 \fadv + 8 \beta \fadv > 0$,
where $\beta \equiv \Pi_{\rm gas} / \Pi$,
and $\fadv \equiv \Qadv/\Qvis$. The dotted and dashed lines both
draw $\delta = 0$, but correspond to $\beta = 2/5$ and $\fadv = 0$,
and to $\fadv = 1/3$ and $\beta = 0$, respectively. The region below
the dotted line has $\delta < 0$ and is for stable SSDs, which are
radiative cooling-dominated ($\fadv \to 0$) and gas
pressure-supported ($\beta > 2/5$). The region between the dotted
and the dashed lines has $\delta > 0$ and is for unstable SSDs,
which are radiation pressure-supported ($\beta < 2/5$) but not yet
advective cooling-dominated ($\fadv < 1/3$). These two regions were
already known in the literature. The region between the dashed
and the solid line has $\delta < 0$ and is obviously for slim disks,
which are advective cooling-dominated ($\fadv > 1/3$) and radiation
pressure-supported ($\beta \to 0$), and are stable.
What is new is that, because of the limitation of $\dmax$,
accretion flows can possibly take the form of slim disks
only in the inner region, i.e., in the region $r < r_{\rm crit}$.

Figure~5 may have interesting implications.
If the accretion rate $\dm$ of an accretion flow is sufficiently
small at large radii (i.e., below the dotted line),
then the flow can behave as an SSD throughout
(here we don't want to discuss the inability of SSD
model for the inner region of a black hole accretion
flow, such as for the flow's transonic motion).
If, however, $\dm$ at large radii is in the unstable region or in
the no-solution region, then it seems that the only possible
way for accretion to proceed is that the flow loses its
matter continuously in the form of outflows.
Such outflows must be so strong that $\dm$ keeps being below the
dotted line all the way till the inner region $r < r_{\rm crit}$.
In the inner region there are two possibilities for the accretion
flow: either $\dm$ is still below the dotted line and the flow keeps
being an SSD, or $\dm$ is in the unstable region. In the latter case
the flow may undergo a limit cycle, i.e., oscillating between
the SSD state and the slim disk state in the vertical axis
direction of Figure~5. Such a limit cycle behavior has been
investigated extensively in the literature
(e.g., Szuszkiewicz \& Miller 2001 and references therein), and
is the likely way for slim disks to be realized.
A related remark is that provided $\dm$ in the outer
region of accretion flows is not sufficiently small,
outflows seem to be unavoidable, as already observed
in many high energy astrophysical systems that are believed
to be powered by black hole accretion.

\section{Discussion}

We have shown that when the gravitational force in the vertical direction
of black hole accretion disks is correctly calculated,
the relationship $\cs/\OmK H =$ constant,
which is valid only for geometrically thin disks,
does not hold for slim disks; and that there exists a maximally
possible mass accretion rate for each radius in the outer
region of optically thick accretion flows, so that only the flow's inner
region can possibly take the form of slim disks, and outflows
from the outer region must be produced
in order for slim disks to be realized. We stress that only
one change has been made in obtaining these results, i.e.,
using the explicit form of the PW potential, instead of
the H\={o}shi form of this potential, to calculate the
vertical gravitational force, while all the assumptions,
equations, and methods for solutions are kept exactly the same as
in the slim disk model (e.g., in the excellent book of Kato et al. 1998).

All our results here are based on a local analysis.
Though the similar local analysis was often used in the literature
(e.g., Abramowicz et al. 1995; Chen et al. 1995; Kato et al. 1998),
it is worth checking our results by constructing global solutions
of original differential equations for black hole accretion flows,
like what was done by, e.g., Chen \& Wang (2004),
but with a revised vertical gravitational force.
We plan to do this in a subsequent work.

A natural question is why we concentrate our attention here only
on slim disks and do not touch another type of black hole
accretion flows, namely optically thin advection-dominated
accretion flows (ADAFs, Narayan \& Yi 1994; Abramowicz et al. 1995),
which are also geometrically not thin.
We think that the ADAF model is also likely to suffer the same
problem in the vertical direction as the slim disk model does,
because the problem is resulted from a purely hydrodynamic consideration
and is related only to the geometrical thickness of the flow.
In particular, the usage of relation $\cs/\OmK H =$ constant
for ADAFs is questionable.
What is different is that for optically thin flows the
radiation processes are more complicated and the vertical
integration is not as easy to perform as for slim disks.
In addition, ADAFs are known to correspond to very low accretion
rates and to have a maximally possible accretion rate
at each radius (e.g., Abramowicz et al. 1995),
so the problem may have no impact on ADAFs in respect of
the accretion rate. However, we still wish to make one more
comment. Many 2-dimensional or 3-dimensional numerical simulations
of viscous radiatively inefficient accretion flows revealed the
existence of convection-dominated accretion flows instead of
ADAFs (e.g., Stone, Pringle, \& Begelman 1999; Igumenshchev \&
Abramowicz 2000; McKinney \& Gammie 2002; Igumenshchev, Narayan,
\& Abramowicz 2003). According to our results here, the 1-dimensional
ADAF model might have hidden inconsistencies in the vertical
direction, so that ADAFs could not be obtained in those
multidimensional numerical simulations.

\acknowledgments

We thank Li Xue for beneficial discussion and the referee for prompt
and helpful comments. This work was supported by the National Science
Foundation of China under Grants No. 10503003 and 10673009.

\clearpage

\begin{figure}
\plotone{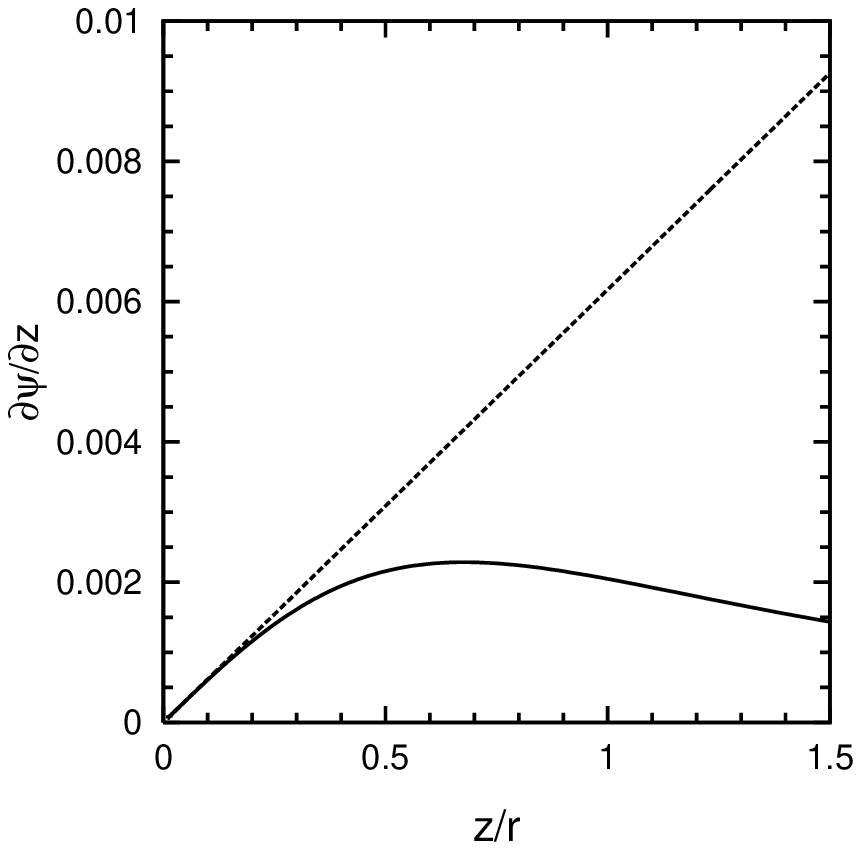}
\caption{Vertical gravitational force $\p \psi / \p z$ for varying
$z/r$ at $r=10\rg$, calculated using the explicit form of
the PW potential (the solid line) and the H\={o}shi form of this potential
(the dashed line), respectively.
\label{fig1}}
\end{figure}

\clearpage

\begin{figure}
\plotone{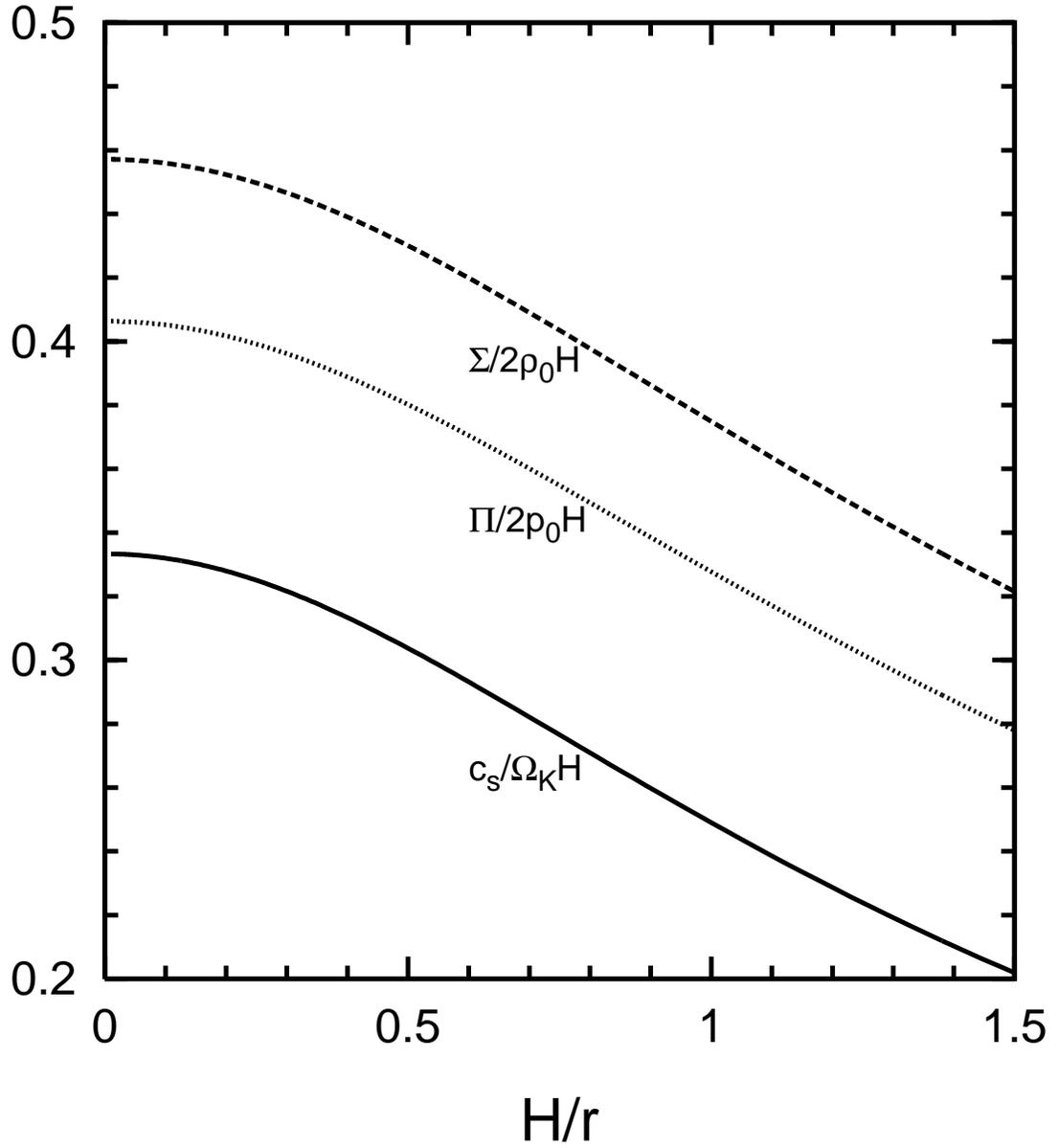}
\caption{
Variations of $\Sigma/2\rho_0 H$ (the dashed line),
$\Pi/2p_0 H$ (the dotted line), and $\cs/\OmK H$ (the solid line)
with $H/r$ at $r=10\rg$.
\label{fig2}}
\end{figure}

\clearpage

\begin{figure}
\plotone{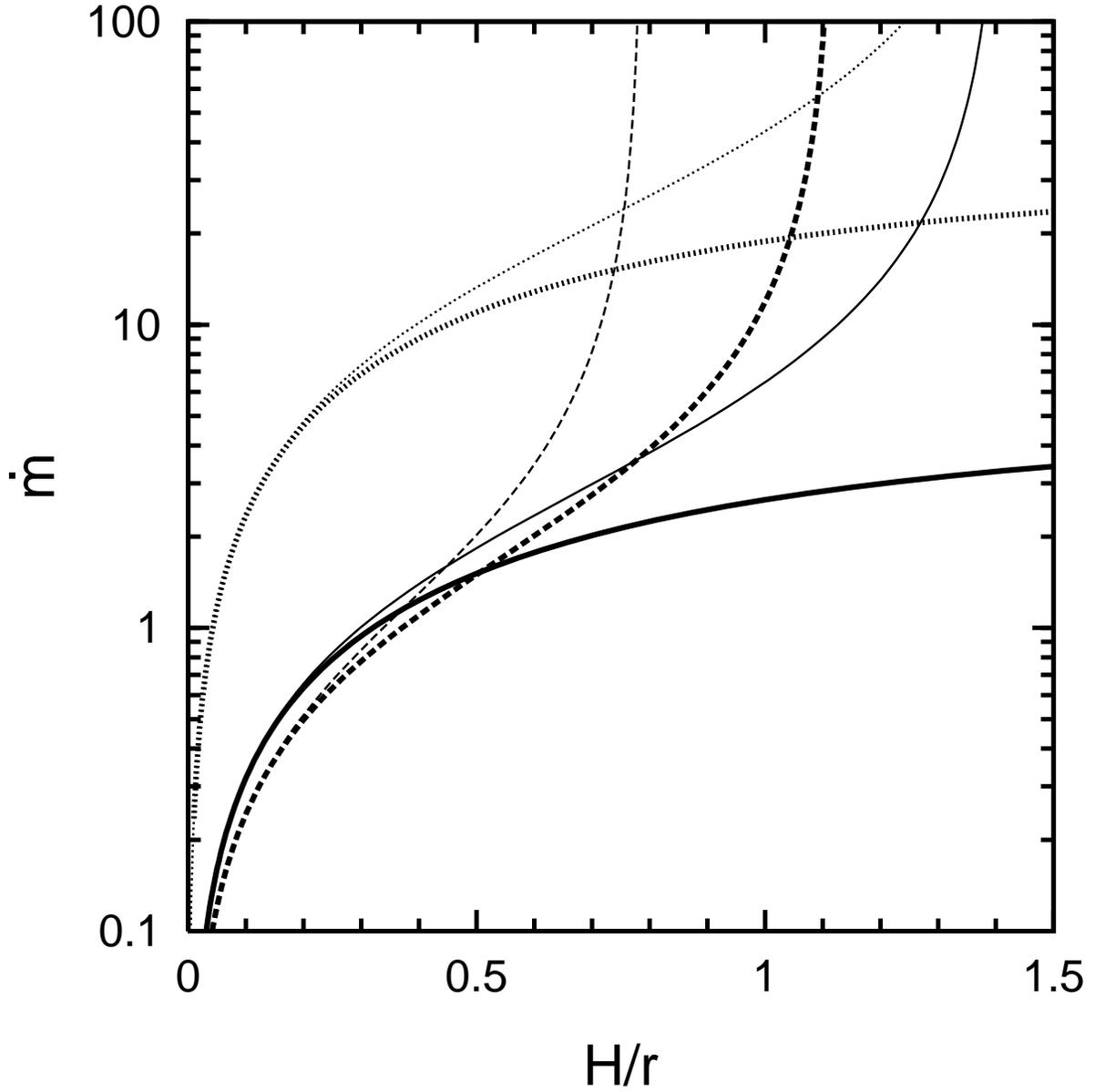}
\caption{
Thermal equilibrium solutions with the H\={o}shi form of the PW potential
(thin lines) and with the explicit PW potential (thick lines)
at $r=5\rg$ (dashed lines), $r=50\rg$ (solid lines), and $r=500\rg$
(dotted lines).
\label{fig3}}
\end{figure}

\clearpage

\begin{figure}
\plottwo{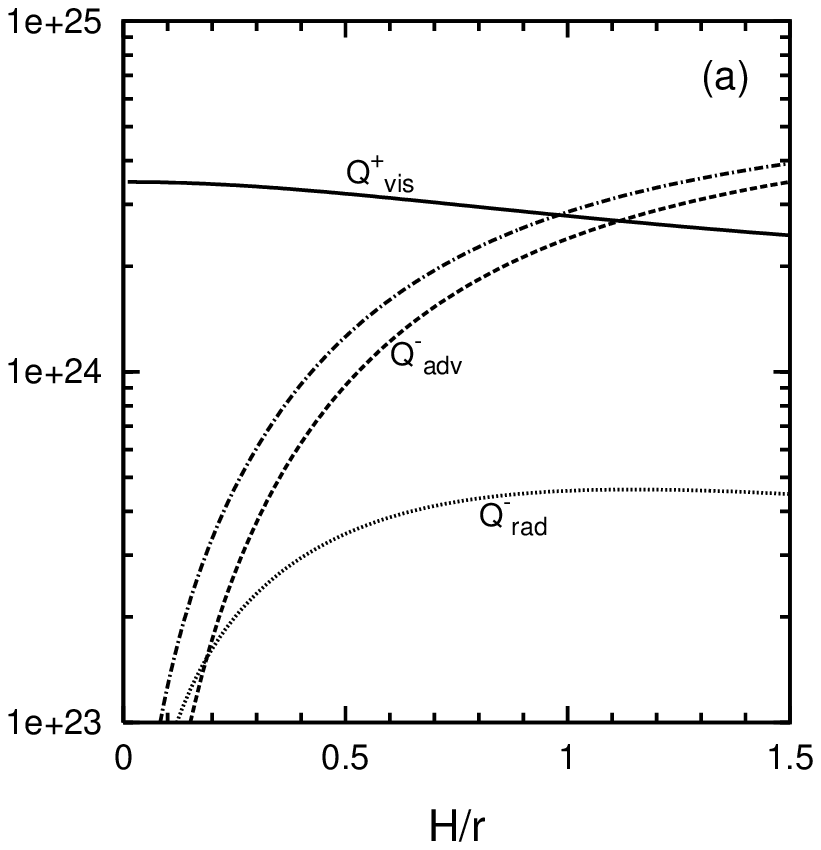}{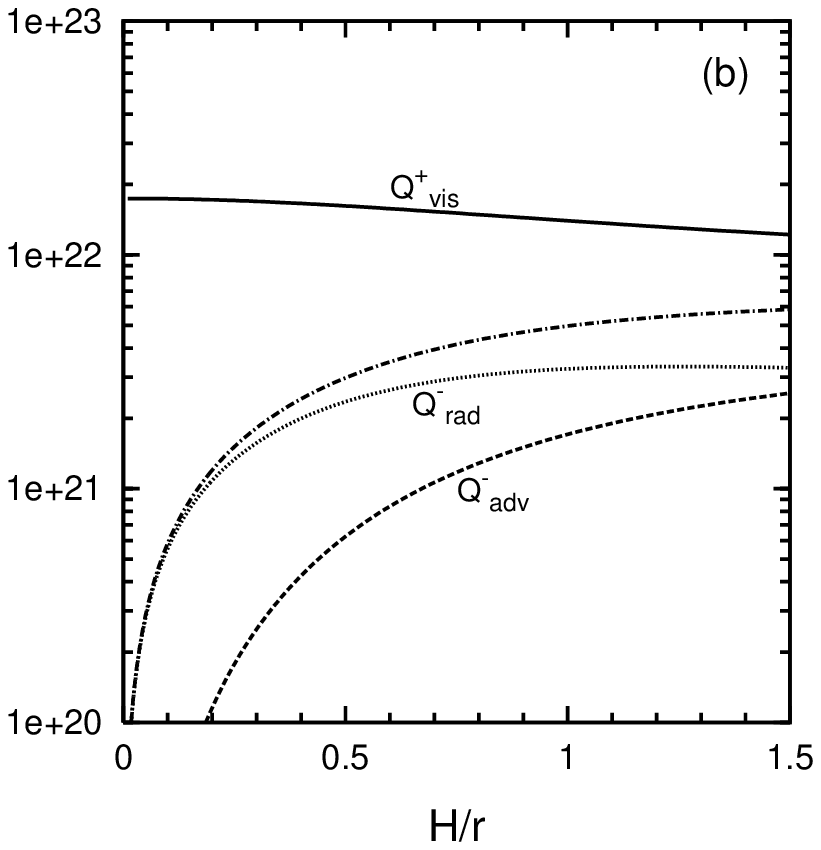}
\caption{Variations of $\Qvis$ (the solid line), $\Qadv$ (the dashed line),
$\Qrad$ (the dotted line), and $\Qadv+\Qrad$
(the dot-dashed line) with $H/r$. (a) is for $\dm=10$ and $r=5\rg$,
(b) is for $\dm=10$ and $r=50\rg$.
Numbers attaching the vertical axis are in units of
erg cm$^{-2}$s$^{-1}$.
\label{fig4}}
\end{figure}

\clearpage

\begin{figure}
\plotone{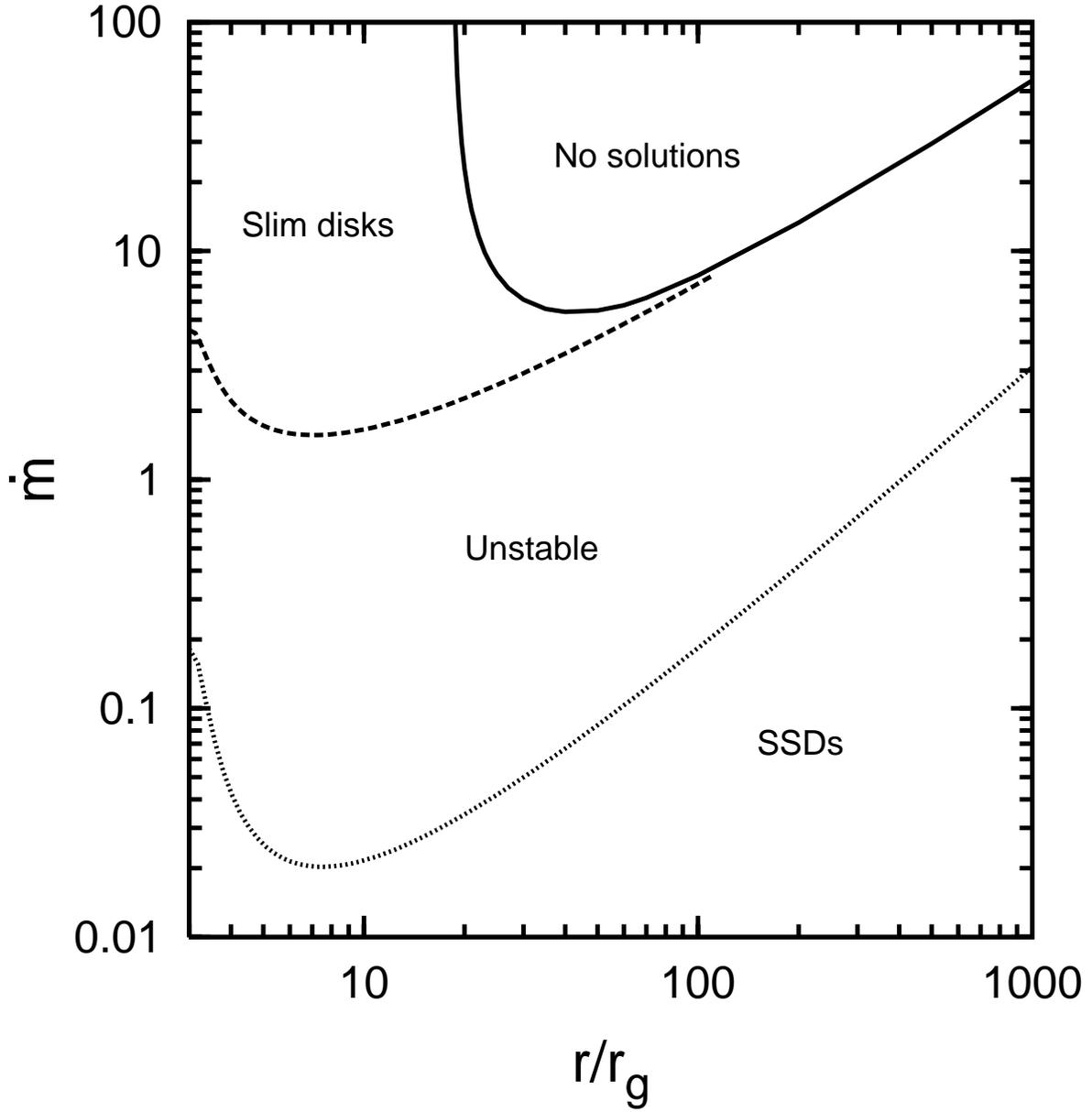}
\caption{Distribution of thermal equilibrium solutions.
The solid line is for $\dmax$, the dashed line
for $\fadv = 1/3$, and the dotted line for $\beta = 2/5$.
\label{fig5}}
\end{figure}

\end{document}